\documentclass[11pt,twoside]{article}
\usepackage{asp2014}

\aspSuppressVolSlug
\resetcounters

\newcommand{\msun}{${\mathrm{M}}_{\odot}$} 
\newcommand{\mjup}{$\rm M_{\rm Jup}$}

\begin{document}

\title{
Detectability of substellar companions\\ around white dwarfs with Gaia}
\author{Roberto~Silvotti,$^1$ Alessandro~Sozzetti,$^1$, Mario~Lattanzi,$^1$
Roberto~Morbidelli$^1$
\affil{$^1$INAF--Osservatorio Astrofisico di Torino, 
Strada dell'Osservatorio 20,\\ 10025 Pino Torinese, Italy; 
\email{silvotti@oato.inaf.it}; \email{sozzetti@oato.inaf.it};
\email{lattanzi@oato.inaf.it}; \email{morbidelli@oato.inaf.it}.}}

\paperauthor{Roberto~Silvotti}{silvotti@oato.inaf.it}
{}{Istituto Nazionale di Astrofisica}{Osservatorio Astrofisico di Torino}
{Pino Torinese}{}{10025}{Italy}
\paperauthor{Alessandro~Sozzetti}{sozzetti@oato.inaf.it}
{}{Istituto Nazionale di Astrofisica}{Osservatorio Astrofisico di Torino}
{Pino Torinese}{}{10025}{Italy}
\paperauthor{Mario~Lattanzi}{lattanzi@oato.inaf.it}
{}{Istituto Nazionale di Astrofisica}{Osservatorio Astrofisico di Torino}
{Pino Torinese}{}{10025}{Italy}
\paperauthor{Roberto~Morbidelli}{morbidelli@oato.inaf.it}
{}{Istituto Nazionale di Astrofisica}{Osservatorio Astrofisico di Torino}
{Pino Torinese}{}{10025}{Italy}

\begin{abstract}
To date not a single-bona fide planet has been identified orbiting a single 
white dwarf.
In fact we are ignorant about the final configuration of $>$95\%
of planetary systems.
Theoretical models predict a gap in the final distribution of orbital periods, due to the opposite effects of stellar mass loss (planets pushed outwards) 
and tidal interactions (planets pushed inwards) during the RGB and the AGB
stellar expansions.
Over its five year primary mission, Gaia is expected to astrometrically detect 
the first (few tens of) WD massive planets/BDs giving first evidence 
that WD planets exist, at least those in wide orbits.
In this article we present preliminary results of our simulations 
of what Gaia should be able to find in this field.
\end{abstract}

\section{Introduction}
%
In recent years, the question of the planet survivability during the
post-MS evolution has attracted some interest and the first models were 
computed (Villaver \& Livio 2007, 2009; Nordhaus et al. 2010; Passy
et al. 2012; Spiegel 2012; Mustill \& Villaver 2012; Nordhaus \& Spiegel 2013;
Villaver et al. 2014).
When the host star experiences the red giant or asymptotic giant expansion,
the interplay between stellar mass loss and tidal effects determine whether
the planetary orbits expand (when stellar mass loss dominates) or shrink
(when tidal effects dominate).
In this game various parameters like the planetary mass, the initial 
semi-major axis and the eccentricity are crucial.
At the end of the evolution, when the star becomes a white dwarf (WD), 
we expect therefore that some planets have migrated outwards and some 
inwards, so that the final distribution of orbital periods shows a gap.
Close to the star we expect to find only massive companions in very 
tight orbits (those for which the initial planetary mass was large enough to survive the common envelope phase), while all the other planets were 
presumably pushed out at several AUs from their host star.

Most of the known planets orbiting Main Sequence (MS) or subgiant stars
have been detected through Radial Velocities (RVs) or transits.
For white dwarfs, the small number of narrow lines limits the precision that 
we can reach with the RV technique.
For what concerns transits, the small WD radii are at the same time and advantage and a limit, implying very deep transits (objects smaller than the Moon could easily be detected), but also very small transit probabilities.
The only sistematic attempt to detect WD planetary transits gave no positive results (Faedi et al. 2011, Braker et al. these proceedings).
In fact only objects very close to the star can be detected from their transit signature and/or from the reflex motion they cause to the star, if these 
objects are massive enough (tens or Jovian masses, see e.g. Maxted et al. 2006).
For what concerns distant planets, those outside the period gap, direct 
imaging is not the solution (see Hogan et al. 2009), being efficient only 
for very young objects: it could work only for 2nd generation planets 
if such planets exist.
Pulsation timing has proved to have serious problems (Dalessio et al. 2013 and these proceedings) but remains interesting at least to set upper limits to the planetary frequency around WDs (Don Winget et al. these proceedings).
Eclipse timing is more reliable (at least in one case, NN Ser, it is giving 
interesting results, Beuermann et al 2013, Marsh et al. 2014 and
these proceedings, Parsons et al. 2014), but is limited only to circumbinary 
planets orbiting eclipsing systems.
The most promising method to detect WD planets in wide orbits is certainly 
the astrometry and the recent successful launch of Gaia suggests that in the 
coming years major improvements in this field will be achieved.
In this context we have started a work to simulate what Gaia can find.

\section{Gaia astrometric detections}

\subsection{Preliminary results}

\articlefigure{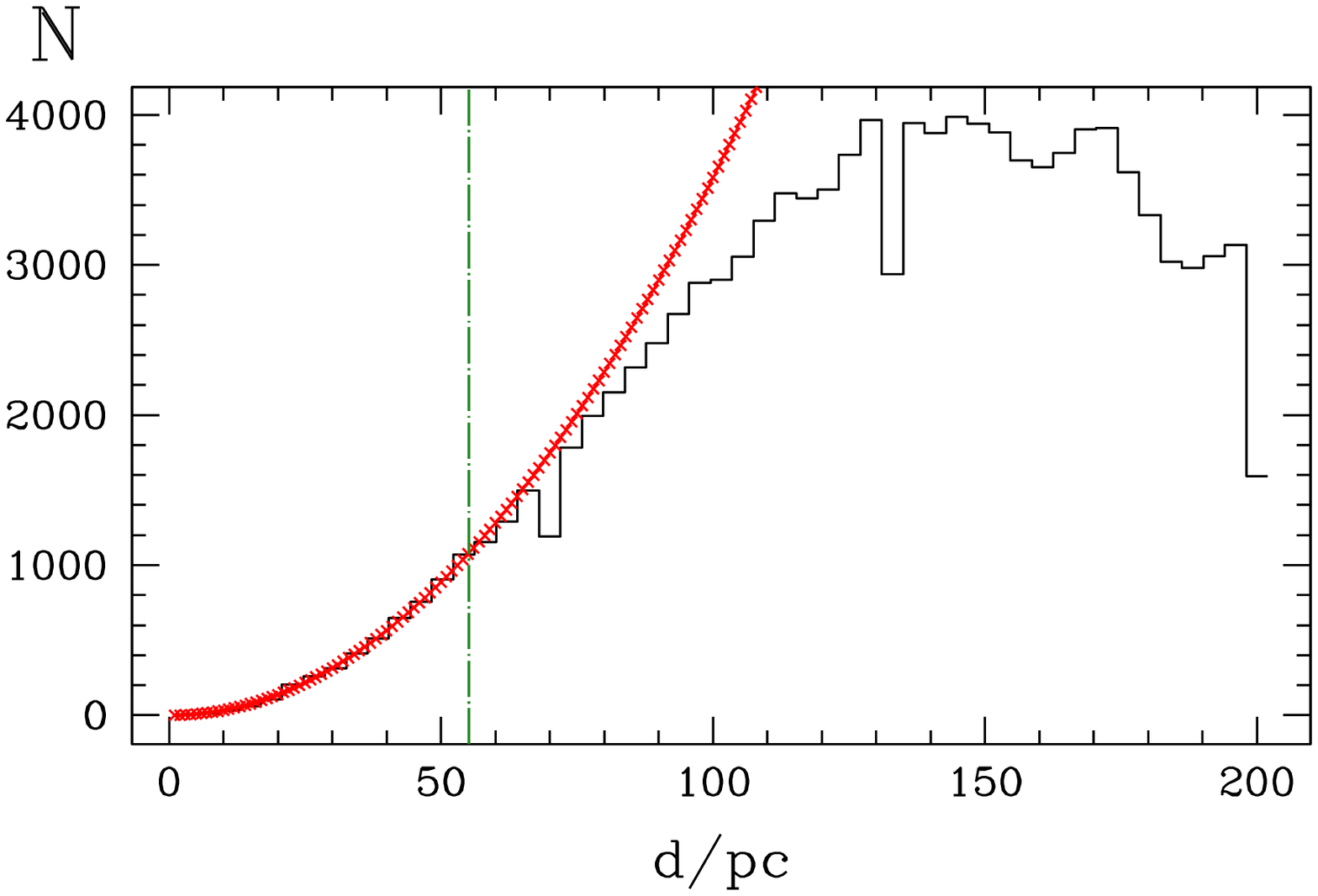}{fig1}{White dwarf distribution as a function of 
distance. The expected trend assuming a constant space density without any cutoff in magnitude is shown (red in the electronic version).
The vertical (green) line represents the completeness limit.}

First we have built up a WD catalogue using the Besan\c{c}on galaxy model 
(Robin et al. 2003)\footnote{\it http://model.obs-besancon.fr/},
including 4 populations (thin and tick disk, spheroid and bulge) and kinematics,
but excluding binaries.
Limiting our catalogue in magnitude (R$<$20.5) and distance (d$<$200 pc),
the total number of objects is 116,295 with a completeness limit near 55 pc
(Fig.~1).
Our simulations are based on the double-blind set-up by Casertano et al. (2008),
considering 5 years of nominal mission and using a pre-launch error model.
We have considered 5 mass ranges for the companion: 1--3, 3--7, 7--13 
and 17--80 \mjup.
Up to 17 \mjup\ the frequency of planets in the MS follows a power law:
$ dN \propto M^{\alpha} P^{\beta} dlnM~dlnP$ with $\alpha=-0.31$ and $\beta=0.26$
(Cumming et al. 2008).
%
From 17 to 80 \mjup\ we have used a uniform distribution with an integral 
frequency of 0.5\% (Ma \& Ge 2014).
We have then considered orbital periods in MS larger then 0.4 years with
all the other orbital elements following a uniform distribution in their full
range of values, except for the eccentricity which varies between 0 and 0.6.
Finally, we have assumed that during the evolution from the MS to the WD cooling 
track, the orbits have expanded by a constant factor 2.5 (corresponding to the 
stellar mass loss of a star with a MS mass of 1.5 \msun).
No tidal effects were considered.
This very simple scheme of orbital expansion is in some way coherent with
an intrinsic limitation of the catalogue obtained from the Besan\c{c}on galaxy 
model, in which the mass of all white dwarfs is assumed to be constant and equal to 0.6 \msun.
These aspects will be improved in the next set of simulations in which a new
catalogue with more realistic WD masses will be used.

The efficiency of the astrometric planet detection for a companion mass of
5, 15 and 50 \mjup\ respectively, is shown in Fig.~2.
In Table~1 we summarize our results.

\articlefigure{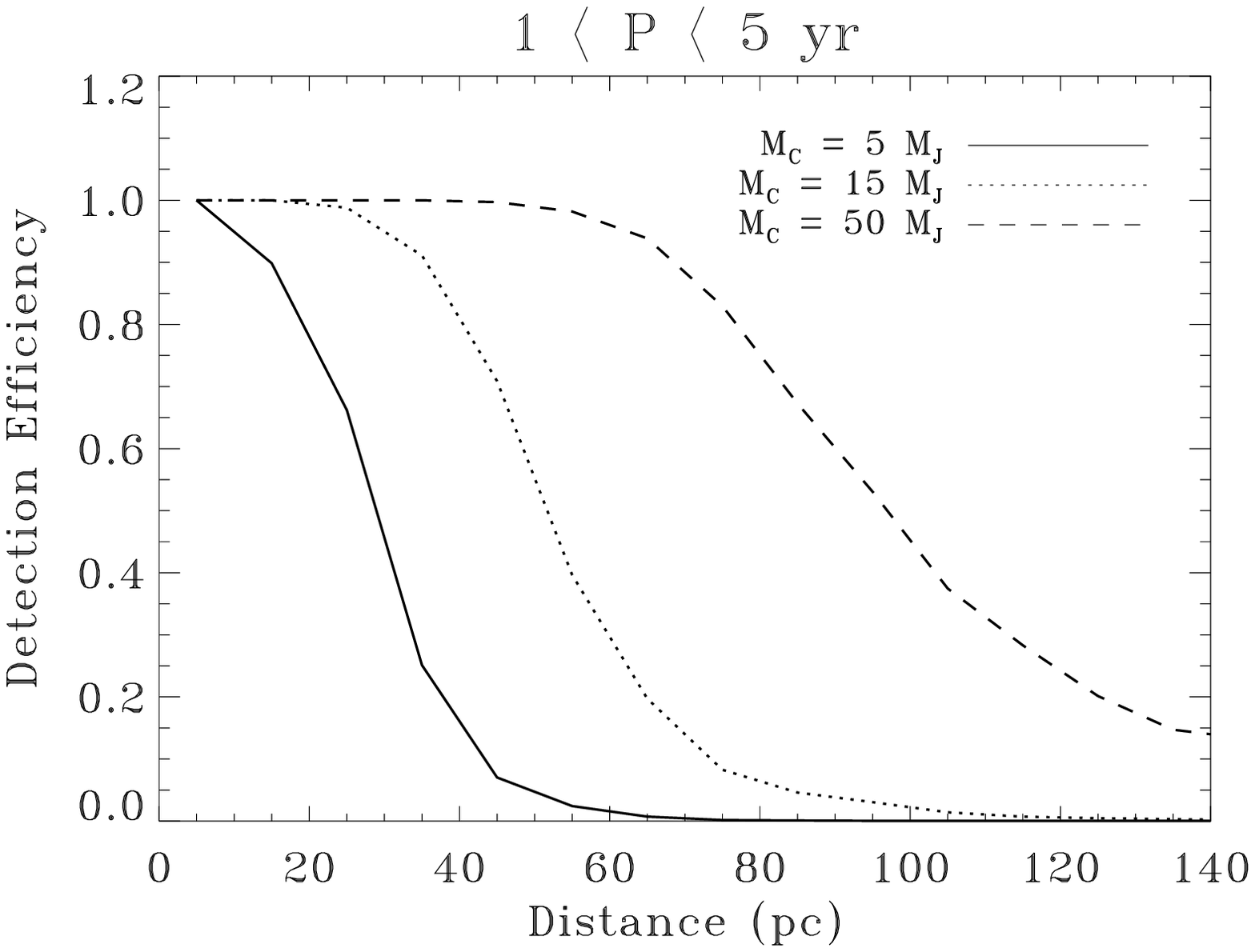}{fig2}{Gaia detection efficiency curves at S/N$>$3,
corresponding to a 99.9\% confidence level (see Sozzetti et al. 2014 for more
details). The typical errors on companion mass recovery are 20-30\%.}


\begin{table} 
\centering
\caption[]{Gaia simulations results}
\vspace{2mm}
\begin{tabular}{|c|c|c|c|c|}
\hline
 & P$_{\rm ORB}$=1-2 yrs & P$_{\rm ORB}$=2-3 yrs & P$_{\rm ORB}$=3-5 yrs &
 TOT (1-5 yrs)\\
\hline
 1-3 \mjup  &   0 &   0 &   0 &   0\\
 3-7 \mjup  &   2 &   1 &   0 &   3\\
 7-13 \mjup &   5 &   4 &   0 &   9\\
13-17 \mjup &  14 &   8 &   4 &  26\\
17-80 \mjup &  71 &  14 &  21 & 106\\
\hline
\end{tabular}
\end{table}

\subsection{Future work}

The results summarized in Table~1 show that Gaia is more sensitive to 
substellar companions with orbital periods between 1 and 
2 years.\footnote{This fact, apparently in contradiction with the 
power-law frequency of MS planets and with the detection efficiency of Gaia, 
which increases at longer periods, is related to the assumed rigid orbital expansion (constant factor of 2.5) caused by the stellar mass loss, which redistributes the orbital periods and pushes the MS periods exceeding 
2 years beyond the upper limit considered of 5 years.}
However, recent theoretical results show that the outer limit of the period gap
is expected to be near 2 yrs (Mustill \& Villaver 2012).
Thus it is likely that Table~1 overestimates the number of planets/BDs with orbital periods shorter than $\sim$2 years.
This problem will be partly solved with a new series of simulations that will
make use of a new catalogue in which more realistic WD and progenitor masses 
will be considered. 
An other improvement that we are considering is to take into account tidal 
effects during the RGB/AGB stellar expansion.




\end{document}